\documentclass[notitlepage,article,showpacs,preprintnumbers,amsmath,amssymb,prb,superscriptaddress,footinbib]{revtex4-1}

\usepackage[utf8]{inputenc}
\usepackage{graphicx}
\usepackage{dcolumn}
\usepackage{bm}
\usepackage{epstopdf}
\usepackage{float}
\usepackage[normalem]{ulem}
\DeclareGraphicsExtensions{.pdf,.eps,.png,.jpg,.mps}

\DeclareUnicodeCharacter{2003}{~}

\usepackage{color, soul}

\usepackage{xcolor}

\begin{document}

\setcounter{page}{1}

\title{Measuring electron spin flip-flops through nuclear spin echo decays}
\author{Evan S. Petersen}
\affiliation{Department of Electrical Engineering, Princeton University}
\author{A. M. Tyryshkin}
\affiliation{Department of Electrical Engineering, Princeton University}

\author{K. M. Itoh}
\affiliation{School of Fundamental Science and Technology, Keio University}

\author{H. Riemann}
\affiliation{Leibniz-Institut für Kristallzüchtung}
\author{N. V. Abrosimov}
\affiliation{Leibniz-Institut für Kristallzüchtung}
\author{P. Becker}
\affiliation{PTB Braunschweig}
\author{H.-J. Pohl}
\affiliation{VITCON Projectconsult GmbH}

\author{M. L. W. Thewalt}
\affiliation{Department of Physics, Simon Fraser University}

\author{S. A. Lyon}
 \affiliation{Department of Electrical Engineering, Princeton University}

\begin{abstract}
We use the nuclear spin coherence of $^{31}$P donors in $^{28}$Si to determine flip-flop rates of donor electron spins. Isotopically purified $^{28}$Si crystals minimize the number of $^{29}$Si flip-flops, and measurements at 1.7~K suppress electron spin relaxation. The crystals have donor concentrations ranging from $1.2\times10^{14}$ to $3.3\times10^{15}~\text{P/cm}^3$, allowing us to detect how electron flip-flop rates change with donor density. We also simulate how electron spin flip-flops can cause nuclear spin decoherence. We find that when these flip-flops are the primary cause of decoherence, Hahn echo decays have a stretched exponential form. For our two higher donor density crystals ($> 10^{15}~\text{P/cm}^3$), there is excellent agreement between simulations and experiments. In lower density crystals ($< 10^{15}~\text{P/cm}^3$), there is no longer agreement between simulations and experiments, suggesting a different, unknown mechanism is limiting nuclear spin coherence. The nuclear spin coherence in the lowest density crystal ($1.2 \times 10^{14}~\text{P/cm}^3$) allows us to place upper bounds on the magnitude of noise sources in bulk crystals such as electric field fluctuations that may degrade silicon quantum devices.
\end{abstract}

{\global\let\newpage\relax\maketitle}

\section{Introduction}

Quantum devices utilizing spins, electron spin resonance (ESR) and nuclear magnetic resonance (NMR) in solids, and magnetic resonance imaging (MRI) can all be affected by electron spin flip-flops. This decoherence mechanism causes errors in spin-based quantum devices,\cite{Tyryshkin2011} affects electron and nuclear coherence times in ESR and NMR, and can be utilized for dynamic nuclear polarization\cite{Wollan1976} to enhance signal strength in MRI.\cite{Dementyev2008} A flip-flop occurs when two dipole-dipole coupled electron spins, one spin up and the other spin down, swap their spin states. This is possible so long as the combined energy of the spins does not change. The rates of flip-flops (spin diffusion) can be calculated in idealized scenarios,\cite{Bloembergen1949, Nolden1996} but in practice inhomogeneous broadening from other mechanisms can obscure their calculation or measurement.\cite{Kuhns1987, Nolden1996, Tyryshkin2011, Blank2016} For flip-flops of $^{31}$P donor electron spins in silicon, measurements of their rates have previously required the use of magnetic field gradients\cite{Tyryshkin2011, Blank2016} in combination with multiple sets of Hahn echo decays. Here we demonstrate how Electron-Nuclear Double Resonance (ENDOR) experiments may be used to determine electron spin flip-flop rates without the use of such gradients. We also demonstrate how to combine these experiments with simulations to determine local ($\sim1~\mu$m scale) inhomogeneous linewidths within crystals and find that they are narrower than global inhomogeneity across the sample volumes.

We measure electron spin flip-flop rates using nuclear Hahn echo decays in $^{28}$Si crystals with $^{31}$P donor concentrations ranging from $1.2\times10^{14}$ to $3.3\times10^{15}~\text{P/cm}^3$. We show that our experiments are not impacted by other known, previously measured decoherence mechanisms. \cite{Morton2008, Abe2010, Petersen2016} We then compare these results to simulations using a model of nuclear spins experiencing decoherence from electron spin flip-flops. Our comparison yields excellent agreement for our crystals with the two highest donor densities. Decoherence in the two lower density crystals is evidently dominated by other, unknown mechanisms. One possible mechanism is electric field fluctuations, and we can set an upper bound on the magnitude of the field fluctuations in our bulk silicon crystals.

\section{Experimental Methods}

Experiments are carried out in a $\sim$0.335 Tesla magnetic field. Each measurement begins by applying a preparation sequence of microwave and rf pulses that creates a coherence on the electron spins of the donors, and then transfers this coherence from the electron spins to the nuclear spins.\cite{Morton2008} The nuclear spins evolve freely for a time $\tau$, are refocused by an rf $\pi$ pulse, and then evolve for another time $\tau$, creating the usual Hahn echo. The nuclear coherence is transformed back into an electron coherence, where it can be measured in an electron spin echo following a microwave $\pi$ pulse. Magnitude detection of the spin echo was then used to remove phase noise caused by fluctuations in the applied magnetic field.\cite{Tyryshkin2003} A light-emitting diode (1050~nm) was flashed for 20~ms following each experiment to thermalize spins between measurements.\cite{Tyryshkin2011}

Other known nuclear spin decoherence mechanisms, specifically 1) $^{29}$Si flip-flops and 2) electron spin relaxation (T$_{1e}$), have negligible impact on these experiments. For the concentration of $^{29}$Si in our crystals (50ppm), $^{29}$Si flip-flops  are expected to decohere $^{31}$P donor nuclear spins in $\sim$100~s.\cite{Petersen2016} At the temperature we perform experiments (1.7~K), T$_{1e}$ is of the order of hours.\cite{Morton2008} Neither process is important on the few second timescale of our experiments.

\section{Experimental Results}

Measurements of nuclear echo decays for each of the crystals are plotted in blue in Fig. 1. More heavily doped samples have faster echo decays than lightly doped samples, and the shape of the echo decays also varies with donor concentration. We first fit (red dashes) the echo decays to the function:\cite{Anderson1962, deSousa2003, Mims1968}
\begin{eqnarray}
v(\tau)=\exp\left(-\left(\frac{2\tau}{T_{\textrm{2}}}\right)^n\right)
\label{eq3}
\end{eqnarray}

\noindent where $2\tau$ is the total time of the Hahn echo experiment, $T_2$ describes the time to decay to a value of 1/e, and $n$ is a stretch factor describing the shape of the decay. This function is often associated with various spectral diffusion decoherence mechanisms, though here we use it because it is convenient for fitting our decay data. In addition, simulations of Hahn echo decays using our flip-flop model, described in Section IV, are plotted as green diamonds in Fig. 1. The dependence of $n$ on phosphorus density as extracted from the fits is plotted as red squares in Fig. 2, which starts at $\sim$~0.6 at the highest donor density and increases to $\sim$~1.1 as the donor density decreases. We also fit the simulated decays using Eq. 1 and consistently find $n$ $\sim$~0.6 for all densities (plotted in green diamonds in Fig. 2).

\begin{figure}
\centering
\includegraphics[scale=.6,keepaspectratio]{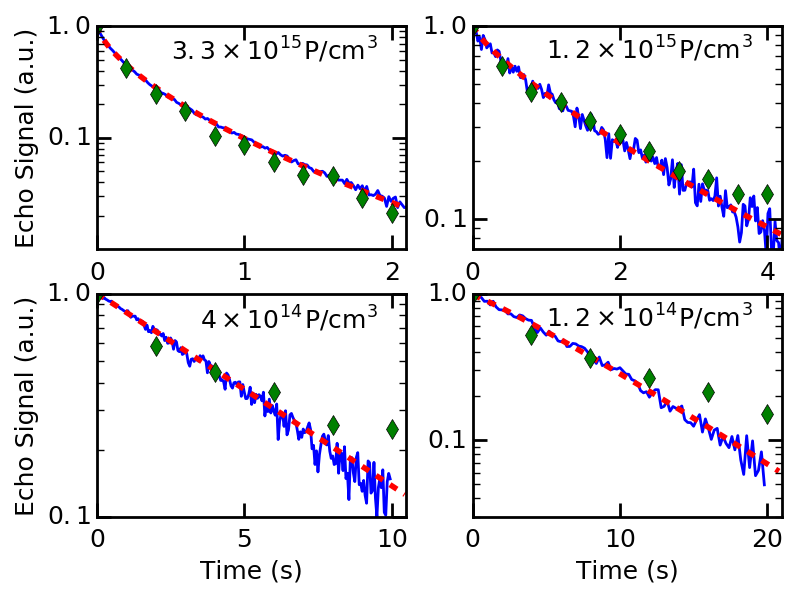}
\caption{\label{fig:decays} (color online) Nuclear spin Hahn echo decays for neutral $^{31}$P donors in isotopically enriched $^{28}$Si crystals with varying donor densities at 1.7~K with magnetic field ($\sim~$0.335~T) oriented along [001]. Experimental decays are plotted in solid blue, fits to Eq. 1 in dashed red, and stochastic model decays (described in Section \ref{sec:theory}) are plotted in green diamonds. The donor concentration of each sample is given for each plot.}
\end{figure}

\begin{figure}
\centering
\includegraphics[scale=.5,keepaspectratio]{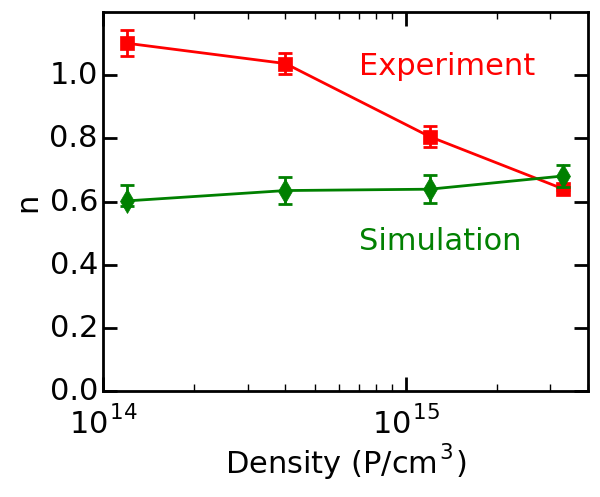}
\caption{\label{fig:fits} (color online) Donor concentration dependence of stretch factor, $n$, derived from the fits of Eq. 1 to experimental decays (red squares) and simulated stochastic model decays (green diamonds) of nuclear spin echoes for $^{31}$P donors in silicon at 1.7~K. Where not shown, error bars fall within markers. Lines are guides to the eye.}
\end{figure}

\section{Theory}
\label{sec:theory}

Here we describe the physics behind our results, first showing how nuclear spin coherence is affected by electron spin flip-flops between a pair of donors and then extending that example to a large-scale model of many donors in a uniformly doped crystal. The Hamiltonian for a pair of donors, each with spin-1/2 electron and nuclear spins, is given by:
\begin{eqnarray}
\begin{split}
&\mathcal{H} =\mathcal{H}_{Z}+\mathcal{H}_{A}+\mathcal{H}_{dd}\\*
&\mathcal{H}_{Z} = \hbar(\omega_{e,i}S_{Z_i}-\omega_{n,i}I_{Z_i}+\omega_{e,j}S_{Z_j}-\omega_{n,j}I_{Z_j})\\*
&\mathcal{H}_{A} = \hbar(\omega_{A,i}S_{Z_i}I_{Z_i}+\omega_{A,j}S_{Z_j}I_{Z_j})\\*
&\mathcal{H}_{dd} = \hbar\omega_{dd}(2S_{Z_i}S_{Z_j} - \frac{1}{2}\left(S_i^+S_j^- + S_i^-S_j^+\right))
\end{split}
\label{eq}
\end{eqnarray}

\noindent where $\mathcal{H}_Z$ contains the Zeeman terms, $\mathcal{H}_A$ contains the hyperfine interactions of the donor electrons and their nuclear spins, $\mathcal{H}_{dd}$ is the electron dipole-dipole interaction, $S_{Z_i}$ and $I_{Z_i}$ are the spin-1/2 operators for the electron and nuclear spins, respectively, for donor $i$, $\omega_{e,i}$ and $\omega_{n,i}$ are the Zeeman frequencies of donor $i$'s electron and nuclear spins, $\omega_{A,i}$ is the hyperfine coupling between the nuclear and electron spins of donor $i$, $\omega_{dd}$ is the dipole-dipole coupling between the electron spins of the two donors, and $S_i^+$ and $S_i^-$ are the raising and lowering operators for the electron spin of donor $i$.
 
For two donors with the same nuclear spin state but opposite electron spin orientations, the two electron spins can swap states without changing the energy of the system. The term $S_i^+S_j^- + S_i^-S_j^+$ in the electron spin dipole-dipole interaction $\mathcal{H}_{dd}$ leads to these flip-flops. The dipole-dipole coupling factor, $\omega_{dd}$, is given by:
\begin{eqnarray}
\omega_{dd}=\frac{\zeta\left(1-3cos^2\left(\theta\right)\right)}{2r^3}
\label{eq2}
\end{eqnarray}

\noindent where $\zeta=\gamma_i\gamma_j\mu_0\hbar/4\pi$, $\gamma_i$ and $\gamma_j$ are the gyromagnetic ratios of two electrons, $\mu_0$ the vacuum permeability, $\hbar$ the reduced Planck constant, $r$ the distance between donors in the pair, and $\theta$ the angle of the vector connecting the donors relative to the global magnetic field, B$_0$.

The evolution of a pair of donors is key to understanding our experiments. As a specific example we consider a pair of donors after our preparation pulses are applied, with one donor in the nuclear spin superposition state $ \frac{1}{\sqrt{2}}\left(|\mathord{\uparrow \Uparrow}\rangle+i|\mathord{\uparrow \Downarrow}\rangle\right)$ and the other donor in the pure state  $|\mathord{\downarrow\Downarrow}\rangle$ (neglecting an inconsequential phase from the preparation pulses). Here we use the notation of spin $\uparrow$ and $\downarrow$ for electron spin states and $\Uparrow$ and $\Downarrow$ for nuclear spin states. We write the combined state of the donors using the convention $|$donor~1,~donor~2$\rangle$ as $\psi_0 =  \frac{1}{\sqrt{2}}\left(|\mathord{\uparrow \Uparrow}, \mathord{\downarrow \Downarrow} \rangle+i|\mathord{\uparrow \Downarrow}, \mathord{\downarrow \Downarrow} \rangle\right)$. For simplicity we assume both donors have the same hyperfine coupling ($\omega_{A,i}=\omega_{A,j}$) and nuclear spin Zeeman frequencies ($\omega_{n,i}=\omega_{n,j}$). In the absence of dipole-dipole interactions ($\omega_{dd}=0$), the Hahn echo experiment evolves the initial $\psi_0$ state to $\psi_{\tau}$ after time $\tau$, then to $\psi_{\tau\pi}$ after a $\pi$ rotation on the nuclear spins for donors with electron spin $\uparrow$, and then finally to $\psi_{\tau\pi\tau}$ following another time $\tau$ when an echo signal is formed. These states are:
\begin{subequations}
\begin{align}
&\psi_0 =  \frac{1}{\sqrt{2}}\Big(|\mathord{\uparrow\Uparrow,\downarrow\Downarrow}\rangle + i|\mathord{\uparrow\Downarrow,\downarrow\Downarrow}\rangle \Big)\\*
\xrightarrow{\tau}~&\psi_{\tau} =   \frac{1}{\sqrt{2}}e^{i \tau ( \delta\omega_e - \omega_A)/2}\Big(|\mathord{\uparrow\Uparrow,\downarrow\Downarrow}\rangle + i e^{i \tau(\omega_{Z_n}+\omega_A/2)} |\mathord{\uparrow\Downarrow,\downarrow\Downarrow}\rangle\Big)\\*
\xrightarrow{\pi}~&\psi_{\tau\pi} =  -\frac{1}{\sqrt{2}}e^{i \tau ( \delta\omega_e - \omega_A)/2}\Big(|\mathord{\uparrow\Downarrow,\downarrow\Downarrow}\rangle + i e^{i \tau(\omega_{Z_n}+\omega_A/2)} |\mathord{\uparrow\Uparrow,\downarrow\Downarrow}\rangle\Big)\\*
\xrightarrow{\tau}~&\psi_{\tau\pi\tau} =  -\bigg[\!\!\bigg[e^{i \tau (\omega_{Z_n} + \delta\omega_e -\omega_A/2)}\bigg]\!\!\bigg]\left(\frac{1}{\sqrt{2}}\Big(|\mathord{\uparrow\Downarrow,\downarrow\Downarrow}\rangle + i|\mathord{\uparrow\Uparrow,\downarrow\Downarrow}\rangle \Big)\right)= C \psi_0^*
\end{align}
\label{eq:noddEv}
\end{subequations}

\noindent where $\delta\omega_e = \omega_{e,i}-\omega_{e,j}$ is the difference in electron Zeeman frequencies between the two donors, the double brackets highlight an inconsequential global phase $C$, and $\psi_0^*$ is the transformation of $\psi_0$ due to the nuclear spin $\pi$ rotation. Thus, at the end of the sequence nuclear coherence is fully recovered.

The evolution is more complicated when $\omega_{dd}\neq 0$. During the first $\tau$ period, one half of the pair state ($|\mathord{\uparrow\Uparrow,\downarrow\Downarrow}\rangle$) accumulates a global phase while the other half ($|\mathord{\uparrow\Downarrow,\downarrow\Downarrow}\rangle$) is involved in an electron spin flip-flop, resulting in the state $\psi_{\tau}^{dd}$:
\begin{eqnarray}
\psi_0 \xrightarrow{\tau} \psi_{\tau}^{dd} =  \frac{1}{\sqrt{2}} e^{i \tau (\omega_{dd} +\delta\omega_e - \omega_A)/2} \Big(|\mathord{\uparrow\Uparrow,\downarrow\Downarrow}\rangle + i e^{i \tau (\omega_{Z_n}+\omega_A/2)}e^{-i \tau \delta\omega_e /2}(p_1|\mathord{\uparrow\Downarrow,\downarrow\Downarrow}\rangle + p_2 |\mathord{\downarrow\Downarrow,\uparrow\Downarrow}\rangle) \Big)
\label{eq:firstFreeEv}
\end{eqnarray}
 where amplitudes $p_1$ and $p_2$ of the flip-flopping states are defined as:
\begin{eqnarray}
\begin{split}
&p_1 =\cos{\left(\frac{\tau}{2}\sqrt{\omega_{dd}^2+\delta\omega_e^2}\right)} + \frac{i \delta\omega_e}{\sqrt{\omega_{dd}^2+\delta\omega_e^2}}\sin{\left(\frac{\tau}{2}\sqrt{\omega_{dd}^2+\delta\omega_e^2}\right)}\\*
&p_2 = \frac{i \omega_{dd}}{\sqrt{\omega_{dd}^2+\delta\omega_e^2}}\sin{\left(\frac{\tau}{2} \sqrt{\omega_{dd}^2+\delta\omega_e^2}\right)}
\end{split}
\end{eqnarray}

\noindent A subsequent nuclear $\pi$ rotation does not refocus $p_1$ and $p_2$. Instead, it interconverts the donor pair states such that the half involved in flip-flops in Eq. \ref{eq:firstFreeEv}  can no longer flip-flop and the other half now can flip-flop. The resulting states $\psi_{\tau\pi}^{dd}$ and $\psi_{\tau\pi\tau}^{dd}$ are:
\begin{subequations}
\begin{align}
&\psi_{\tau}^{dd}\xrightarrow{\pi}~\psi_{\tau\pi}^{dd} = -\frac{1}{\sqrt{2}} e^{i \tau (\omega_{dd} +\delta\omega_e - \omega_A)/2} \Big(|\mathord{\uparrow\Downarrow,\downarrow\Downarrow}\rangle + i e^{i \tau (\omega_{Z_n}+\omega_A/2)}e^{-i \tau \delta\omega_e /2}(p_1|\mathord{\uparrow\Uparrow,\downarrow\Downarrow}\rangle + p_2 |\mathord{\downarrow\Downarrow,\uparrow\Uparrow}\rangle) \Big)\\
\begin{split}
\\&\xrightarrow{\tau}~\psi_{\tau\pi\tau}^{dd} = -\Bigg[\!\!\Bigg[e^{i \tau (\omega_{Z_n} + \delta\omega_e/2 -\omega_A/2)}\Bigg]\!\!\Bigg]\Bigg(p_1 \frac{1}{\sqrt{2}} \Big(|\mathord{\uparrow\Downarrow,\downarrow\Downarrow}\rangle + i |\mathord{\uparrow\Uparrow,\downarrow\Downarrow}\rangle\Big) +  p_2 \frac{1}{\sqrt{2}}\Big(|\mathord{\downarrow\Downarrow,\uparrow\Downarrow}\rangle + i e^{-i \tau \delta\omega_e} |\mathord{\downarrow\Downarrow,\uparrow\Uparrow}\rangle \Big) \Bigg)\\
&\quad{  }=C'\Bigg(p_1 \psi_0^* + p_2 \psi_{ff}\Bigg)
\end{split}
\end{align}
\label{eq:finalFlipFlopEcho}
\end{subequations}

\noindent where double brackets in $\psi_{\tau\pi\tau}^{dd}$ again indicate an inconsequential global phase $C'$ and $\psi_{ff}$ is a new donor pair state $\frac{1}{\sqrt{2}}\left(|\mathord{\downarrow\Downarrow,\uparrow\Downarrow}\rangle + i e^{-i \tau \delta\omega_e} |\mathord{\downarrow\Downarrow,\uparrow\Uparrow}\rangle \right)$ resulting from flip-flops. At the end of the pulse sequence, the donor pair state $\psi_{\tau\pi\tau}^{dd}$ consists of two parts: one with nuclear coherence recovered in the original state $\psi_0^*$ with probability $|p_1|^2$ and the other with nuclear coherence transferred to the second donor by flip-flops ($\psi_{ff}$) with probability $|p_2|^2$. The nuclear coherence that moves to the second donor accumulates an additional phase $e^{-i \tau \delta\omega_e}$. In ensemble experiments inhomogeneous broadening results in a broad distribution of the detuning $\delta\omega_e$ between donors and therefore a broad distribution of the added phase $e^{-i \tau \delta\omega_e}$. When averaged over an ensemble this distribution results in irreversible decay of the collective nuclear coherence of flip-flopping donor pairs.

The above example (Eqs.~5-7) illustrates the complexity of nuclear spin coherence evolution in the case of an isolated donor pair. The situation is even more complicated in randomly doped crystals where isolated pairs are rare and each donor can simultaneously interact with many other donors. Nuclear spin coherence initially localized on one donor may spread to many donors through flip-flops during the free evolution periods before and after the $\pi$ rotation. This results in irreversible diffusion of spin coherence to other donors and irreversible spin decoherence from the random phases $e^{-i \tau \delta\omega_e}$ acquired on those other donors.

The evolution of a random multi-donor system is intractable, so we need a different approach. As demonstrated above, any flip-flops a donor's electron spin participates in will irreversibly destroy that donor's nuclear spin coherence. Therefore when calculating nuclear spin coherence it is sufficient to track whether a donor's electron spin engages in a flip-flop event rather than trying to monitor the phase of every spin in a crystal. Bloembergen\cite{Bloembergen1949} demonstrated a procedure to model such flip-flop events semiclassically with $\Gamma_{i,j}$, the probability per unit time that electron spins $i$ and $j$ flip-flop. This rate is given by:
\begin{eqnarray}
\Gamma_{i,j} = \left(\frac{2\pi}{\hbar^2}\right)\Bigg|\langle j|\Big[\hbar\omega_{dd}\frac{1}{2}\left(S_1^+S_2^-+S_1^-S_2^+\right)\Big]|i\rangle\Bigg|^2 f(0)
\label{eq:ffrate}
\end{eqnarray}

\noindent where $f(0)$ is the value at 0~Hz of the distribution $f(\delta\omega_e)$ of spin detuning between electron spins in donor pairs. This detuning distribution arises from inhomogeneous broadening of Zeeman frequencies of individual spins in donor ensembles. Inhomogeneous broadening can be caused by magnetic fields from defects with magnetic moments, random strain effects in the crystal, and inhomogeneities in B$_0$. In our calculations we assume a Gaussian shape for $f(\delta\omega_e)$. The free induction decay (FID) experiments measured in our samples suggest a Gaussian inhomogeneous broadening of individual spins with linewidth $\Delta\omega=180$~kHz. This FID-derived broadening measured on a macroscopic scale (sample size $\sim5$~mm) sets an upper bound for the unknown width of $f(\delta\omega_e)$ on a microscopic scale (e.g. 67~nm, the average distance between donors at $3.3\times 10^{15} ~\text{P/cm}^3$). In our simulations the linewidth, $\Delta\omega$, is the only fitting parameter, with the only constraint that $\Delta\omega < 180$~kHz. This parameter inversely controls the rate of flip-flops in Eq. 8. Lastly, in Eq. 8 we use $f(\delta\omega_e)$ at 0~Hz detuning frequency recognizing the fact the $\omega_{dd}$ ($\sim170$~Hz at 67~nm) is much smaller than the width ($\sqrt{2}\Delta\omega$) of $f(\delta\omega_e)$.

Our simulation starts with a number of donors (typically 70 were enough for convergence) randomly distributed in a volume. The size of the volume is set by the donor density of the crystal to simulate. Although our preparation pulse sequence is intended to create a nuclear coherence on a subset of donors in a crystal, the sequence also affects other donors. The four thermal equilibrium donor states $|\mathord{\downarrow \Downarrow} \rangle$, $|\mathord{\uparrow \Downarrow}\rangle$, $|\mathord{\downarrow \Uparrow}\rangle$, and $|\mathord{\uparrow \Uparrow} \rangle$ become the four states $(1/\sqrt{2})(|\mathord{\uparrow \Uparrow} \rangle+i|\mathord{\uparrow \Downarrow} \rangle)$, $(1/\sqrt{2})(|\mathord{\uparrow \Uparrow}\rangle-i|\mathord{\uparrow \Downarrow}\rangle)$, $|\mathord{\downarrow \Uparrow} \rangle$, and $-|\mathord{\downarrow \Downarrow}\rangle$, respectively. Only the first two states have the intended nuclear spin coherence for our experiments. The other two states participate in flip-flops that can decohere the first two states as described above. At our temperature (1.7~K), each of the states occurs with essentially equal probability, and so the donors in our model each start in one of these states at random.

As shown in Eqs. 5-7, during any given free evolution only part of the nuclear spin superposition may be capable of flip-flops. Therefore we must separately record the probabilities for each donor's electron and nuclear spin states at all times, representing each donor in our simulation with four variables: one for each spin state's probability. Visualizations for spin flip-flops in this representation for three donors are given in Fig. 3. An electron spin flip-flop is represented by an exchange of probabilities between variables, shown between the left and middle donors in the transition from Figures 3A to 3B, and between the middle and right donors in Figures 3B to 3C. Exchanges may only take place between variables corresponding to matching nuclear spin states and opposite electron spin states. All other probabilities for the states of a pair of donors are unchanged following a flip-flop. For a pair of variables where one is smaller than the other, the value of the larger variable is reduced by the smaller one, and the smaller variable has its value changed to zero as demonstrated in the transition from Figures 3A to 3B.

\begin{figure}
\centering
\includegraphics[scale=.4,keepaspectratio]{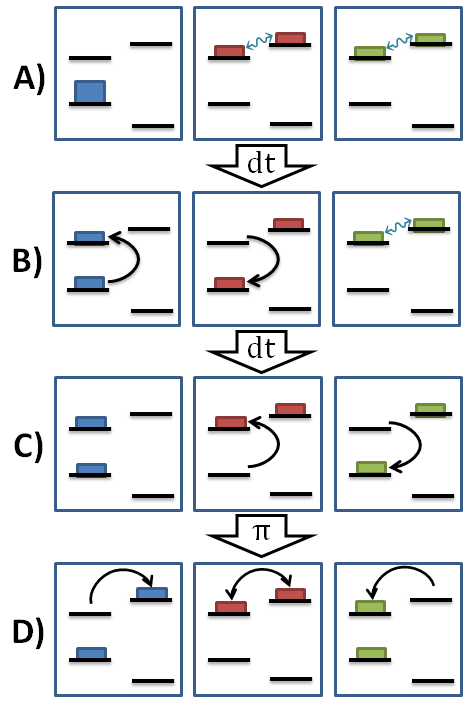}
\caption{\label{fig:flipFlops} (color online) Hahn echo simulation timeline for three donors. A) The red and green donors are each in a coherent superposition state (marked with blue wavy arrows), and the blue donor is in the $|\mathord{\downarrow \Downarrow}\rangle$ state. B) After time $dt$, an electron spin flip-flop can occur between the blue and red donors based on the probability $\Gamma_{i,j}dt$ derived using Eq. \ref{eq:ffrate}, destroying the coherence of the red donor. Only part of the red donor's state was capable of electron spin flip-flops with the blue donor, the rest is unchanged. C) After another time $dt$, an electron spin flip-flop occurs between the red and green donors, decohering the green donor. Even though the probabilities of the energy levels in the red donor are back to how they were in A, there is no longer any nuclear spin coherence. D) A nuclear $\pi$ pulse swaps the nuclear spin polarizations for states with electron spin up in each donor.}
\end{figure}

Free evolution is modeled as a series of time steps, each representing changes in the donor bath occurring within time $dt$. For each donor $j$, the neighboring donor $i$ with the largest $\Gamma_{i,j}$ (most likely to flip-flop with) is first identified and may randomly flip-flop with donor $j$ with probability $\Gamma_{i,j} dt$. If no flip-flop occurs, the next most likely donor $i'$ may randomly flip-flop with donor $j$ with probability $\Gamma_{i',j} dt$. This iteration continues until a flip-flop occurs involving donor $j$ or no neighboring donors remain. The entire procedure is repeated for every donor $j$ in the bath with the conditions that 1) no donor may participate in more than one flip-flop within this time $dt$, 2) there are no double-counting of donor pairs, and 3) no donors can flip-flop if their nuclear spin states do not match.

We assume no errors in our pulses, and model $\pi$ pulses on nuclear spins by swapping the probabilities of states with electron spin up for each donor. After free evolution for a number of time steps corresponding to a total time $\tau$, a $\pi$ pulse applied to the nuclear spins, and another free evolution for time $\tau$, the donors initially in the $(1/\sqrt{2})(|\mathord{\uparrow \Uparrow} \rangle+i|\mathord{\uparrow \Downarrow} \rangle)$ state that have not engaged in electron spin flip-flops produce a Hahn echo. Of these donors, only the center-most spin is counted. By averaging over many iterations and varying $\tau$, we obtain simulated nuclear echo decays.

In the above procedure we only account for decoherence from so-called direct flip-flops\cite{Kurshev1992} (involving a central donor $j$ and one neighbor donor $i$), ignoring any effect from indirect flip-flops (involving two neighbor donors $i$ and $i'$). Indirect flip-flops\cite{Kurshev1992} have little effect on nuclear spins.\cite{deSousa2003, Witzel2012} These indirect processes were previously shown to limit the central donor's electron spin coherence to $\sim1$~s in a lightly doped ($1.2\times10^{14}~\text{P/cm}^3$) sample.\cite{Tyryshkin2011} However, the gyromagnetic ratio of a donor's nuclear spin is three orders of magnitude smaller than the gyromagnetic ratio of the electron spin, so these indirect flip-flops can only limit nuclear spin coherence in our crystals on timescales of 100's to 1000's of seconds. All nuclear $T_2$'s measured in this work were under 10 seconds and therefore must be limited by processes other than indirect flip-flops.

\section{Discussion}

Simulated echo decays from our model in section \ref{sec:theory} are shown as green diamonds in Fig. 1 alongside the measured decays for each crystal. Our simulations produce stretched exponential echo decays, with n = 0.6 at all donor densities, as shown in Fig. 2. The inhomogeneous $\Delta\omega$ widths used in these simulations, from highest to lowest density crystals, are 105~kHz, 70~kHz, 35~kHz, and 5~kHz. Agreement between model and experiment is reasonable for the two higher density samples, but poor for the two lower density samples. The consistent 0.6 stretch factor in modeled decays but not in the experimental data suggests refitting the data to a form appropriate for two independent processes: 
\begin{eqnarray}
v(\tau) = \exp{\left(-2\tau/T_{2a} - \left(2\tau/T_{2ff}\right)^{0.6}\right)}
\label{eq:newFit}
\end{eqnarray}

\noindent where $T_{2ff}$ is the decoherence time from direct flip-flops, and $T_{2a}$ is the decoherence time from other (unknown) interactions. The $T_{2a}$ term introduces a simple exponential decay of the echo. $T_{2ff}$ and $T_{2a}$ times extracted from fits to this form are plotted in Fig. 4. The $T_{2ff}$ times are consistent with an inverse squared donor density dependence ($T_{2ff}\sim 1/\left[^{31}\text{P}\right]^2$) expected from Eq. 8 when $f(0)$ and $\Delta\omega$ are independent of donor concentration. Our model can match the $T_{2ff}$ decay times for the two higher density crystals by using $\Delta\omega=100$~kHz for each crystal, consistent with this assumption. 

\begin{figure}
\centering
\includegraphics[scale=.53,keepaspectratio]{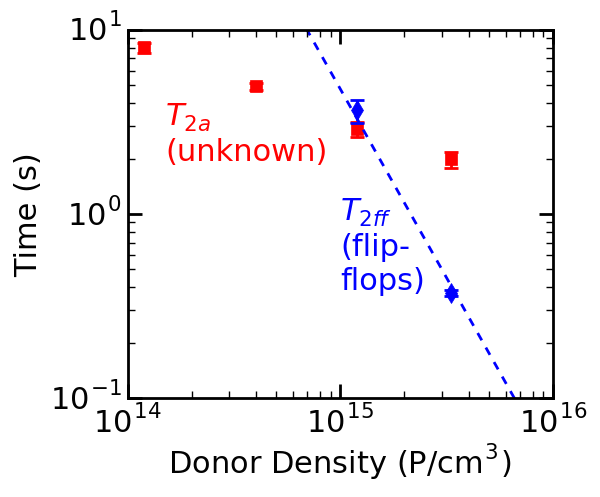}
\caption{\label{fig:newfit} (color online) Donor concentration dependence of $T_{2a}$ (red squares) and $T_{2ff}$ (blue diamonds) extracted from fits of measured echo decays using Eq. \ref{eq:newFit} assuming two decoherence mechanisms. The blue dotted line is an inverse density squared $\left(1/\left[^{31}\text{P}\right]^2\right)$ dependence.}
\end{figure}

 The flip-flop rates in our two higher density crystals, calculated as $1/T_{2ff}$, are $\sim$2.7~Hz and $\sim$270~mHz for the $3.3 \times 10^{15}~\text{P/cm}^3$ and $1.2 \times 10^{15}~\text{P/cm}^3$ crystals, respectively. In comparison, our lowest density sample ($1.2 \times 10^{14}~\text{P/cm}^3$) was previously estimated to have a 1.3~Hz flip-flop rate based on electron spin echo decays with and without an applied field gradient.\cite{Tyryshkin2011} The extracted $T_{2a}$ from fitting Eq. 9 to the nuclear echo decay of that crystal suggests that those previous electron spin measurements may have been influenced by another decoherence mechanism, casting doubt on the 1.3~Hz estimate. Assuming a squared density dependence of $1/T_{2ff}$ based on Eq. 8 and the same $\Delta\omega$ in our highest and lowest density crystals, we now predict a 3~mHz flip-flop rate for the $1.2 \times 10^{14}~\text{P/cm}^3$ sample. Different $\Delta\omega$ widths will inversely scale this predicted rate. 
 
 For very small $\Delta\omega$, which occurs in the limit of broadening only from dipole-dipole interactions between donors (for $1.2\times 10^{14}~\text{P/cm}^3$, a $\Delta\omega\sim66$~Hz wide Lorentzian\cite{Kittel1953}), the approximations in Eq. 8 break down. In this case electron spin flip-flop rates approach $\omega_{dd}$ as $\Delta\omega \rightarrow 0$; approximately 6~Hz for $1.2\times 10^{14}~\text{P/cm}^3$ and 170~Hz for $3.3\times 10^{15}~\text{P/cm}^3$. Recently, a 12.3~Hz rate was obtained by Dikarov \textit{et al.}\cite{Blank2016} using a pulsed field gradient approach at 10~K in a $^{28}$Si epilayer with $10^{16}~\text{P/cm}^3$. Dikarov's sample had a higher donor density than in our crystals (increasing the flip-flop rate), but the strain broadening in epilayers is larger (decreasing the flip-flop rate), so it is difficult to make a quantitative comparison. Our ENDOR method benefits from experimental simplicity, requiring only one echo decay measurement and no accurate external field gradients, but it is limited to samples in which flip-flops are the dominant decoherence mechanism (i.e. $T_{2a} \gtrapprox T_{2ff}$).

 For the highest density sample ($3.3 \times 10^{15}~\text{P/cm}^3$), where direct flip-flops are the dominant source of decoherence, our model reveals a \textit{local} ($\sim 1~\mu$m scale, since flip-flops at larger distances are too slow) electron Zeeman frequency distribution with $\Delta\omega=100$~kHz. In comparison, we can extract a 180~kHz inhomogeneous linewidth from the free induction decay of this sample. The free induction decay measures spins throughout the entire crystal, so it is not surprising that fitting it yields a wider distribution than our model predicts. This wider \textit{global} distribution can differ from local frequency distributions due to variations in strain or defect density as well as long-range field inhomogeneity. We also considered how $^{29}$Si atoms\cite{Abe2010, Witzel2012} and local field inhomogeneity ($\sim$100~nm scale) in our magnet could broaden electron spin resonance lines, but neither of these mechanisms contribute more than $\sim2$~kHz to the linewidth. 

The fitted $T_{2a}$ times exhibit an approximate inverse square root density dependence ($T_{2a}\sim 1/\left[^{31}\text{P}\right]^{0.5}$). We currently do not know the origin of $T_{2a}$, but it dominates the decays of the lower density crystals in Fig. 1. In the two lowest density crystals, $T_{2a}$ prevents us from explicitly determining $T_{2ff}$, and in the $1.2 \times 10^{15} ~\text{P/cm}^3$ crystal $T_{2a}$ is comparable to $T_{2ff}$. One possible source of $T_{2a}$ is electric field noise causing fluctuating Stark shifts of the donor nuclear spins. Fluctuating electric fields could arise from donor-acceptor pair recombination, since impurities are neutralized by the optical pulses used to thermalize spins between measurements.

We can use the nuclear $T_{2a}$ in our lowest density sample to place an upper bound on other possible decoherence mechanisms within the crystal. This bound can be determined by treating the Hahn echo experiment as a filter function and passing a given noise spectrum through it.\cite{deSousa2009} The purely exponential decay we measure suggests using a white noise distribution with amplitude $\nu$. For $T_{2a}=8.2$~s we find $\nu=0.04$~s$^{-1}$. Recent measurements of the hyperfine Stark parameter\cite{Sigillito2015} let us translate this bound into a limit on possible electric fields from charge fluctuations. The linear Stark shift from charge noise is given by\cite{Sigillito2015} $df = \eta_a a E_{\text{strain}} E_{\text{noise}}/\hbar$, where $\eta_a$ is the hyperfine Stark parameter ($-$2.7$\times 10^{-3} \mu$m$^2$/V$^2$),\cite{Sigillito2015} $a$ is the hyperfine coupling constant, $E_{\text{strain}}$ is the equivalent electric field from strain (60~mV/$\mu$m assuming the crystal's 1~kHz ENDOR linewidths come from strain), and $E_{\text{noise}}$ is the electric field from charge noise. We then take $\hbar df$ to be equal to the root-mean-square (RMS) of the noise up to a high frequency cut-off, $\omega_c$, and solve for the corresponding electric field. A recent study on detuning noise in a Si/SiGe singlet-triplet qubit\cite{Qi2017} measured a pink noise spectrum of the form $\alpha_0 / \omega^{0.7}$ with $\alpha_0=47~\text{ns}^{-1.7}$. In that study, $\omega_c$=1/1700~ns which gives an RMS value of $5.7~\mu$eV. Assuming a spacing of 100~nm between quantum dots, the corresponding RMS electric field between the dots is 57~$\mu$V/$\mu$m. For our measured white noise spectrum, the RMS value calculated for the same $\omega_c$ as in the dot study gives an upper limit of the electric field fluctuations of 2~$\mu$V/$\mu$m. The $\sim30\times$ difference between these measurements is not surprising since charge fluctuations near a surface in the singlet-triplet device are expected to be larger than those in the bulk of a high quality silicon crystal.

In conclusion, we demonstrate that donor electron flip-flop rates can be directly measured through donor nuclear spin echo decays. Electron spin flip-flops are a prominent cause of decoherence of neutral $^{31}$P donor nuclear spins in moderately doped ($> 10^{15}~\text{P/cm}^3$) $^{28}$Si crystals. These flip-flops result in the stretched exponential decay of nuclear spin Hahn echoes. The rate of this decay is quadratic in donor density, as expected for electron spin flip-flops with a density-independent linewidth. The local distributions of electron Zeeman frequencies control which electron spins can flip-flop. These local distributions are on the order of 100 kHz wide; about half as wide as distributions of electron spin resonance frequencies measured across entire crystals. In more lightly doped crystals ($< 10^{15}~\text{P/cm}^3$), nuclear spin echoes decay with a simple exponential and rates approximately vary as the square root of donor density. The physical mechanism of the nuclear decoherence in these lowest density $^{28}$Si crystals is not yet known, but an approximate upper bound of $E$~=~2~$\mu$V/$\mu$m can be placed on the RMS value of electric field noise in the lowest density crystal over a frequency range up to $\sim100$~kHz, which is $\sim30\times$ less than the electric field noise recently measured in a Si/SiGe quantum device.\cite{Qi2017}

Work was supported by the NSF and EPSRC through the Materials World Network and NSF MRSEC Programs (Grant No. DMR-1107606, EP/I035536/1, and DMR-01420541), and the ARO (Grant No. W911NF-13-1-0179). M.~L.~W.~T. was supported by the Natural Sciences and Engineering Research Council of Canada (NSERC). The work at Keio has been supported by KAKENHI (S) No. 26220602 and JSPS Core-to-Core Program.

\bibliography{espFull2}
\end{document}